\documentclass[12pt]{article}
\usepackage{authblk}
\usepackage[bookmarksnumbered, colorlinks, plainpages]{hyperref}
\usepackage{amsmath, amsthm, amscd, amsfonts, amssymb, graphicx, color, booktabs, subfig}
\usepackage{lineno}
\numberwithin{equation}{section}
\definecolor{email}{rgb}{0.00,0.00,0.84}
\begin{document}
\setcounter{page}{1}

\title{\large \bf 12th Workshop on the CKM Unitarity Triangle\\ Santiago de Compostela, 18-22 September 2023 \\ \vspace{0.3cm}
\large \bf ATLAS and CMS Collaborations  \\ \vspace{0.1cm}
\LARGE Top-quark spin properties at LHC~\footnote{Copyright [2018] CERN for the benefit of the [ATLAS and CMS Collaborations]. CC-BY-4.0 license.} }

\author{Nello Bruscino\textsuperscript{1} on behalf of the ATLAS and CMS Collaborations \\
        \textsuperscript{1}INFN Sezione di Roma, Rome, Italy \\ }

\maketitle

\begin{abstract}
Due to its high mass top quarks decay before top-flavoured hadrons can be formed. This feature yields experimental access to the top quark polarization and production asymmetries. The large top quark sample moreover enables measurements of other properties, such as the W-boson branching ratios and helicity, and fragmentation functions of the bottom quarks. 
In this contribution, recent measurements of top-quark spin properties at the LHC are presented, including in particular the first evidence of the charge asymmetry in $t\bar{t}$ production (ATLAS), the charge asymmetry in a boosted regime (CMS) and the first measurement of the charge asymmetry in $t\bar{t}+W$ (ATLAS).
\end{abstract} \maketitle

\section{Introduction}
The large mass of the top quark, which is close to the electroweak symmetry breaking scale, indicates that this particle could play a special role in the Standard Model (SM), as well as in beyond the Standard Model (BSM) theories. 
Moreover, the top quark has a very short lifetime ($\tau=0.5 \times 10^{-25}\,\textnormal{s}$) and decays before hadronisation ($\tau_\textnormal{\scriptsize had} \sim 10^{-24}\,\textnormal{s}$) or spin de-correlation take place ($\tau_\textnormal{\scriptsize spin dec.} \sim 10^{-21}\,\textnormal{s}$).
Therefore several properties of the top quark may be measured precisely from its decay products.

Due to the large top-pair production ($t\bar{t}$) cross section for $13\,\textnormal{TeV}$ proton-proton ($pp$) collisions, the Large Hadron Collider (LHC)~\cite{bib:LHC} experiments collect an unprecedented number of top-quark events.
The copious amount of detected events allows for high precision measurements in order to probe predictions of quantum chromodynamics (QCD), which provides the largest contribution to $t\bar{t}$ production. 
This process may also be employed to produce a large, unbiased sample of $W$-bosons and study its properties.

This article focuses on six recent results in the top-quark sector by the ATLAS~\cite{bib:ATLAS} and CMS~\cite{bib:CMS} Collaborations, using proton-proton ($pp$) collisions at LHC.

\section{Measurement of the charge asymmetry in top quark pair production at 13 TeV}
Production of top quark pairs is symmetric at leading-order (LO) under charge conjugation. The asymmetry between the $t$ and $\bar{t}$ originates from interference of the higher-order amplitudes in the $q\bar{q}$ and $qg$ initial states, with the $q\bar{q}$ annihilation contribution dominating. The contribution from electroweak corrections is about 13\% for the inclusive asymmetry.
The $gq \to t\bar{t}q$ production process is also asymmetric, but its cross section is much smaller than $q\bar{q}$. Gluon fusion production is symmetric to all orders.
As a consequence of these asymmetries, the top quark is preferentially produced in the direction of the incoming quark.


A central-forward charge asymmetry for the $t\bar{t}$ production, referred to as the charge asymmetry ($A_\textnormal{\scriptsize C}$), is defined as $A_\textnormal{\scriptsize C}^{t\bar{t}}=\frac{N(\Delta|y|>0)-N(\Delta|y|<0)}{N(\Delta|y|>0)+N(\Delta|y|<0)}$,
where $\Delta|y|=|y(t)|-|y(\bar{t})|$ is the difference between the absolute value of the top-quark rapidity $|y_t|$ and the absolute value of the top-antiquark rapidity $|y_{\bar{t}}|$.
Contributions from new particles can lead to significant modifications of the asymmetry compared to the SM prediction.

The measurement of the $t\bar{t}$ charge asymmetry is performed using data corresponding to an integrated luminosity of $139\,\textnormal{fb}^{-1}$ from the ATLAS experiment~\cite{bib:chargeasymmetry}.
It is performed in the single-lepton channel combining both the resolved and boosted topologies of top quark decays.
A Bayesian unfolding procedure is used to infer the asymmetry at parton level, correcting for detector resolution and acceptance effects.

\begin{figure}[!htb]
\centering
\subfloat[]{\includegraphics[width=0.44\linewidth]{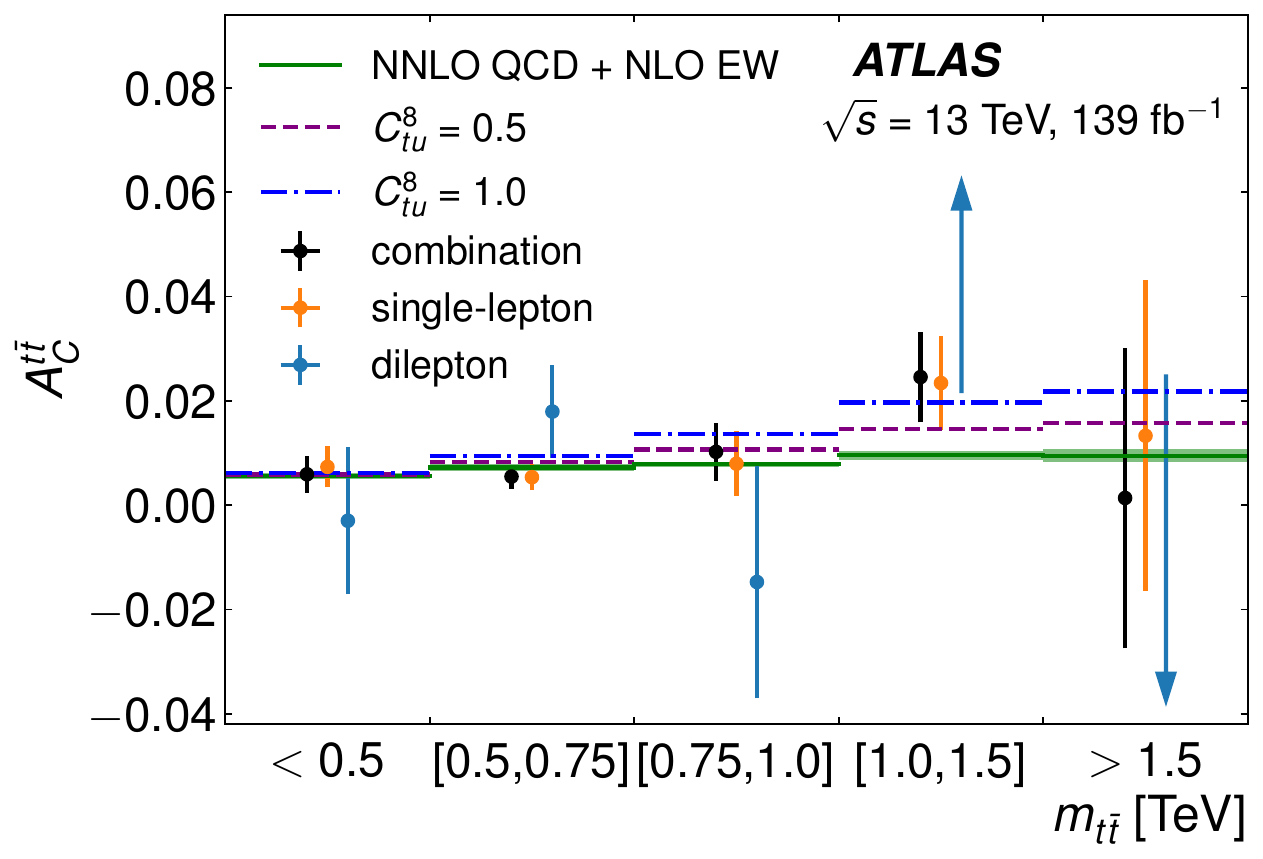}}
\subfloat[]{\includegraphics[width=0.44\linewidth]{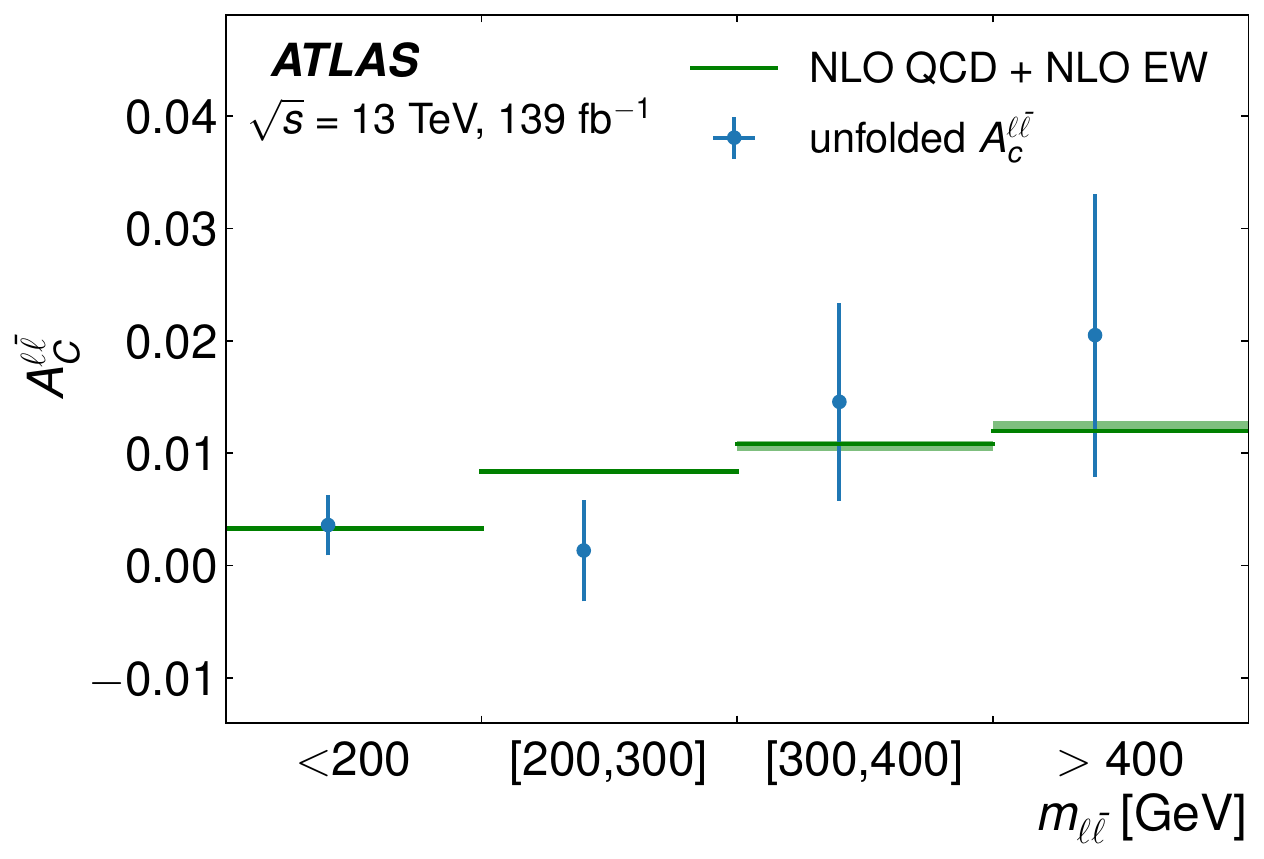}}
\caption{(a)The unfolded differential charge asymmetries as a function of the invariant mass of the reconstructed top-quark pair in data. 
(b) The unfolded differential leptonic asymmetries as a function of the reconstructed lepton pair in dilepton channel data. 
Vertical bars correspond to the total uncertainties. Shaded regions show SM theory predictions.~\cite{bib:chargeasymmetry}}
\label{fig:prop_asymmetry}
\end{figure}

The inclusive $t\bar{t}$ charge asymmetry is measured as $A_\textnormal{\scriptsize C} = 0.68\% \pm 0.15\% \textnormal{(stat+syst.)}$, which differs from zero by 4.7 standard deviations.
It corresponds to the first evidence for charge asymmetry in $pp$ collisions.
Differential measurements are performed as a function of the invariant mass and longitudinal boost of the $t\bar{t}$ system.
Both inclusive and differential measurements are found to be compatible with the SM predictions, at NNLO in perturbation theory with NLO electroweak corrections,
and are shown in Figure~\ref{fig:prop_asymmetry}.

\section{Measurement of the $t\bar{t}$ charge asymmetry in events with highly Lorentz-boosted top quarks at $\sqrt{s}=13$ TeV}
\begin{figure}[!htb]
\centering
\begin{minipage}[c]{0.50\textwidth}
\includegraphics[width=0.95\linewidth]{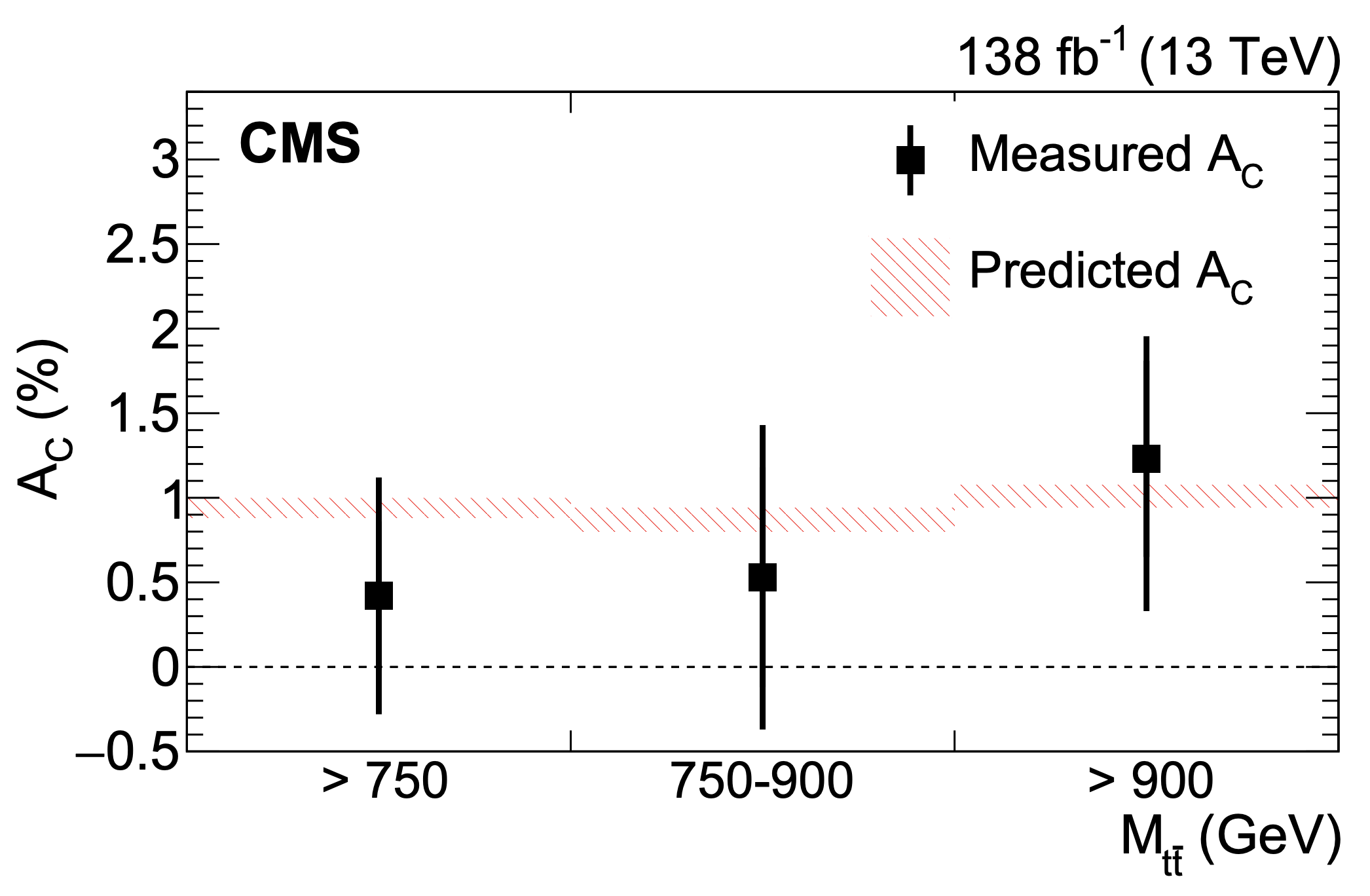}
\end{minipage}
\begin{minipage}[c]{0.49\textwidth}
\caption{Unfolded $A_\textnormal{\scriptsize C}$ in the full phase space presented in different mass regions after combining the $\mu+$jets and $e+$jets channels. 
The vertical bars represent the total uncertainties. The measured values are compared to the theoretical prediction, including NNLO QCD and NLO EW corrections.~\cite{bib:chargeasymmetryboosted}}
\label{fig:asymmetryboosted}
\end{minipage}
\end{figure}

This measurement of the charge asymmetry in top quark pair events focus on highly
Lorentz-boosted top quarks decaying to a single non-isolated lepton and overlapping jets~\cite{bib:chargeasymmetryboosted}.
The analysis is performed using proton-proton collisions at $\sqrt{s}=13$ TeV with the CMS detector corresponding to an integrated luminosity of $138\,\textnormal{fb}^{-1}$ . 
The top quark charge asymmetry is measured for events with a $t\bar{t}$ invariant mass larger than 750 GeV and corrected for detector and acceptance effects using a binned maximum likelihood fit. 
The measured top quark charge asymmetry of $0.42\% + 0.65\% - 0.69\% \textnormal{(stat.+syst.)}$ is in good agreement with the SM prediction at next-to-next-to-leading order in quantum chromodynamic perturbation theory with next-to-leading-order electroweak corrections (NNLO QCD + NLO EW). 
The result is also presented for two invariant mass ranges, $750–900$ and $>900$ GeV, as shown in Figure~\ref{fig:asymmetryboosted}.

\section{Search for leptonic charge asymmetry in $t\bar{t}W$ production in final states with three leptons at $\sqrt{s}=13$ TeV}
The search for the leptonic charge asymmetry ($A_\textnormal{\scriptsize C}^\ell$) of top-quark–antiquark pair production in association with a $W$ boson~\cite{bib:chargeasymmetyttW} is performed using final states
with exactly three charged light leptons (electrons or muons) and is based on $\sqrt{s}=13$ TeV proton–proton collision data collected with the ATLAS detector corresponding to an integrated luminosity of $139\,\textnormal{fb}^{-1}$.
.A profile-likelihood fit to the event yields in multiple regions corresponding to positive and
negative differences between the pseudorapidities of the charged leptons from top-quark and
top-antiquark decays is used to extract the charge asymmetry. 
At reconstruction level, the asymmetry is found to be $-12.3 \pm 13.6 \textnormal{(stat.)} \pm 5.1 \textnormal{(syst.)} \%$. 
An unfolding procedure is applied to convert the result at reconstruction level into a charge-asymmetry value in a fiducial volume at particle level, $-11.2 \pm 17.0 \textnormal{(stat.)} \pm 5.5 \textnormal{(syst.)} \%$. The Standard Model expectations for these two observables are calculated using Monte Carlo simulations with NLO+PS precision in QCD and including NLO EW corrections and in agreement with the measurements.

\section{Measurement of the top quark polarization and $t\bar{t}$ spin correlations using dilepton final states at $\sqrt{s}=13$ TeV}
The measurements of the top quark polarization and $t\bar{t}$ spin correlations uses events containing two oppositely charged leptons produced in proton-proton collisions at a center-of-mass energy of 13 TeV. The data were recorded by the CMS experiment in 2016 and correspond to an integrated luminosity of $35.9\,\textnormal{fb}^{-1}$~\cite{bib:spincorrelation}. 
A set of parton-level normalized differential cross sections, sensitive to each of the independent coefficients of the spin-dependent parts of the $t\bar{t}$  production density matrix, is measured. The measured distributions and extracted coefficients are compared with SM predictions from simulations at NLO accuracy in QCD, and from NLO QCD calculations including EW corrections. 
All measurements are found to be consistent with the expectations of the Standard Model. 
The normalized differential cross sections are used in fits to constrain the anomalous chromomagnetic and chromoelectric dipole moments of the top quark to $-0.24 < C_{tG}/\Lambda^2 < 0.07\,\text{TeV}^{-2}$ and $-0.33 < C^I_{tG}/\Lambda^2 < 0.20\,\text{TeV}^{-2}$, respectively, at the 95\% confidence level.

\section{Measurement of the polarisation of $W$ bosons produced in top-quark decays using dilepton events at $\sqrt{s}=13$ TeV}
As previously said, the $t\bar{t}$ production process can be employed to produce a large and unbiased sample of $W$-bosons and study its properties, like the polarisation. 
Using proton–proton collision data recorded by the ATLAS detector at a centre-of-mass energy of $\sqrt{s}=13$ TeV ($139\,\textnormal{fb}^{-1}$), the measurement~\cite{bib:Whelicity} is performed selecting events decaying into final states with two charged leptons (electrons or muons) and at least two $b$-tagged jets. The polarisation is extracted from the differential cross-section distribution of the $\cos\theta^*$ variable, where $\theta^*$ is the angle between the momentum direction of the charged lepton from the $W$ boson decay and the reversed momentum direction of the $b$-quark from the top-quark decay, both calculated in the W boson rest frame. 
Parton-level results, corrected for the detector acceptance and resolution, show a very good agreement with SM predictions. The measured fractions of longitudinal, left- and right-handed polarisation states are found to be:
$F_0 = 0.684 \pm  0.005 \text{(stat.)} \pm  0.014 \text{(syst.)}$, 
$F_L = 0.318 \pm  0.003 \text{(stat.)} \pm  0.008 \text{(syst.)}$ and 
$F_R = -0.002 \pm  0.002 \text{(stat.)} \pm  0.014 \text{(syst.)}$,
in agreement with the Standard Model prediction.

\section{Measurement of the polarisation of single-top quarks in $t$-channel at $\sqrt{s} = 13\,\textnormal{TeV}$}
The QCD $pp \to t\bar{t}$  process produces unpolarised top quarks because of parity conservation in QCD, while single-top-quark production yields a large sample of highly polarised top quarks and top antiquarks.
The analysis performs a simultaneous measurement of the three components of the top-quark and top-antiquark polarisation vectors in $t$-channel single-top-quark production~\cite{bib:polarisation}.
It is based on data from proton-proton collisions at a centre-of-mass energy of 13 TeV corresponding to an integrated luminosity of $139\,\textnormal{fb}^{-1}$, collected with the ATLAS detector. Selected events contain exactly one isolated electron or muon, large missing transverse momentum and exactly two jets, one being $b$-tagged.
The top-quark and top-antiquark polarisation vectors are measured simultaneously.
In addition, normalised differential cross-sections corrected to a fiducial region at particle level are determined as a function of the charged-lepton angles for top-quark and top-antiquark events separately and inclusively. 

\begin{figure}[!htb]
\centering
\begin{minipage}[c]{0.50\textwidth}
\includegraphics[width=0.75\linewidth]{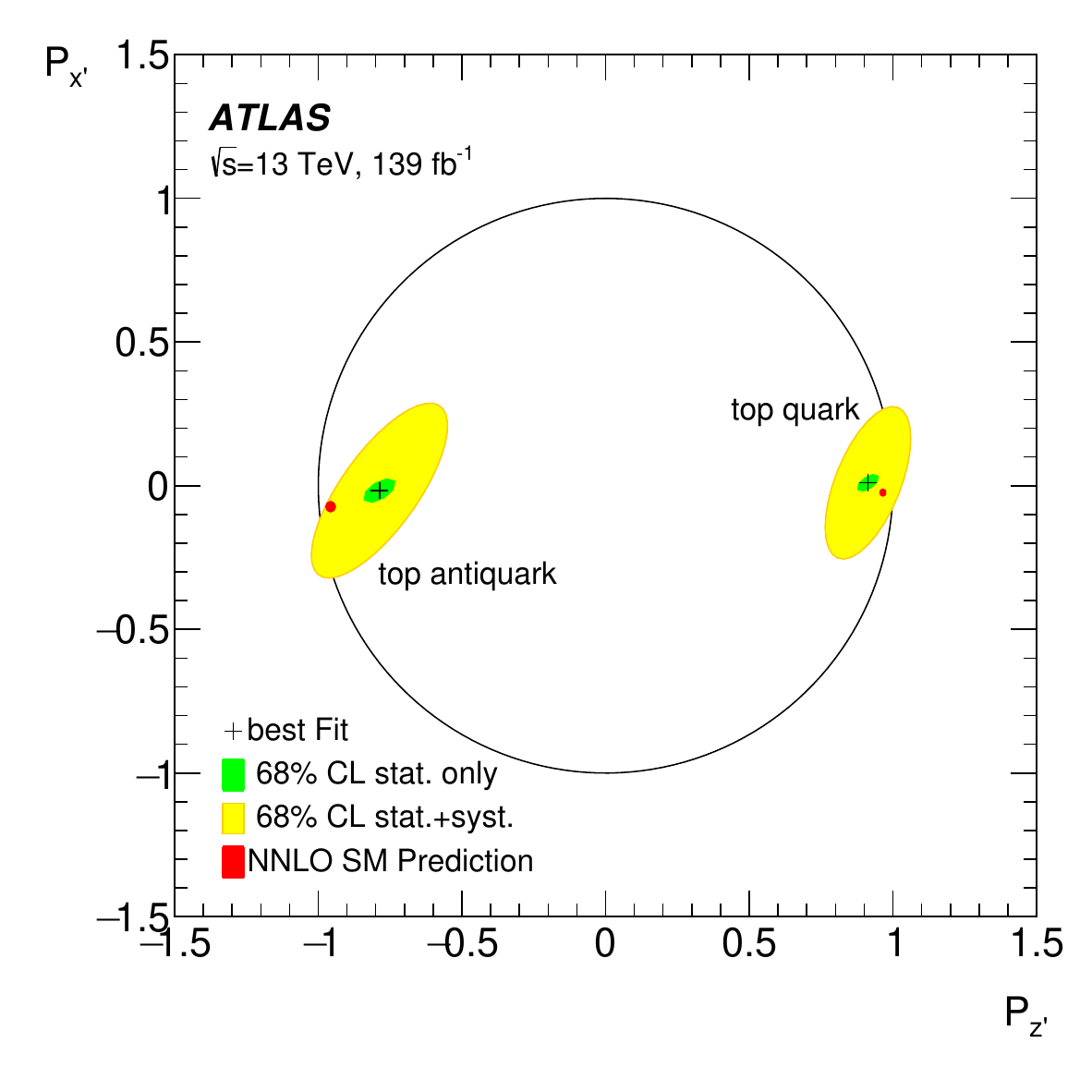}
\end{minipage}
\begin{minipage}[c]{0.39\textwidth}
\caption{(a) Summary of the observed best-fit polarisation measurements ($P_{z'}$, $P_{x'}$) with their statistical-only (green) and statistical+systematic (yellow) contours at 68\% CL.~\cite{bib:polarisation}. }
\label{fig:polarisation}
\end{minipage}
\end{figure}

The three components of the polarisation vector for the selected top-quark event sample are measured to be $P_{x'} = 0.01 \pm 0.18$, $P_{y'} = -0.029 \pm 0.027$, $P_{z'} = 0.91 \pm 0.10$ and for the top-antiquark event sample they are $P_{x'} = -0.02 \pm 0.20$, $P_{y'} = -0.007 \pm 0.051$, $P_{z'} = 0.79 \pm 0.16$.
They are all in very good agreement with the SM predictions at  NNLO accuracy, as shown in Figure~\ref{fig:polarisation}.
The normalised differential cross-sections are in agreement with SM predictions. The derived boundaries on the complex Wilson coefficient of the dimension-six $O_{tW}$ operator are $C_{tW} \in [-0.9,1.4]$ and $C_{itW} \in [-0.8,0.2]$, both at 95\% confidence level; they are also compatible with the SM.

\bibliographystyle{amsplain}

\begin{thebibliography}{99}

\bibitem{bib:LHC} 
L. Evans and P. Bryant (editors), 
2008 JINST 3 S08001.

\bibitem{bib:ATLAS} 
ATLAS Collaboration, 
JINST 3 (2008) S08003.

\bibitem{bib:CMS} 
CMS Collaboration, 
JINST 3 (2008) S08004.

\bibitem{bib:chargeasymmetry}
ATLAS Collaboration, 
JHEP 08 (2023) 077.

\bibitem{bib:chargeasymmetryboosted}
CMS Collaboration, 
Phys. Lett. B 846 (2023) 137703.

\bibitem{bib:chargeasymmetyttW}
ATLAS Collaboration, 
JHEP 07 (2023) 033.

\bibitem{bib:spincorrelation}
CMS Collaboration, 
Phys. Rev. D 100 (2019) 072002.

\bibitem{bib:Whelicity}
ATLAS Collaboration, 
Phys. Lett. B 843 (2023) 137829.

\bibitem{bib:polarisation}
ATLAS Collaboration, 
JHEP11 (2022) 040.

\end{thebibliography}

\end{document}